\documentclass[11pt]{article}
\usepackage{Blois,epsfig}

\def\Journal#1#2#3#4{{#1} {\bf #2}, #3 (#4)}

\def\NPB{{\em Nucl. Phys.} B}
\def\PLB{{\em Phys. Lett.}  B}
\def\PRL{\em Phys. Rev. Lett.}
\def\PRD{{\em Phys. Rev.} D}
\def\YF{\em Yad. Fiz.}
\def\PAN{\em Phys. Atom. Nucl.}
\def\MPLA{{\em Mod. Phys. Lett.} A}
\def\CPC{\em Comp. Phys. Comm.}

\def\be{\begin{equation}}
\def\ee{\end{equation}}
\def\bea{\begin{eqnarray}}
\def\eea{\end{eqnarray}}

\begin{document}
\vspace*{2cm}
\begin{center}
\Large{\textbf{XIth International Conference on\\ Elastic and 
Diffractive Scattering\\ Ch\^{a}teau de Blois, France, May 15 - 20, 
2005}}
\end{center}

\vspace*{2cm}
\title{HIGGS PRODUCTION VIA GLUON FUSION WITH $K_T$ FACTORIZATION}

\author{ A.V. LIPATOV, \underline{N.P. ZOTOV} }

\address{SINP, Moscow State University,\\
119992 Moscow, Russia}

\maketitle\abstracts{We consider the Higgs boson production at high 
energy hadron colliders
in the framework of the $k_T$-factorization approach.
The attention is focused on the dominant gluon-gluon fusion subprocess.
We show that $k_T$-factorization gives a possibility to
investigate the associated Higgs boson and jets production.}

\section{Introduction}
The search for the Higgs boson takes important
part at the Fermilab Tevatron experiments and will be one of the main 
fields of study at the CERN LHC collider. In QCD
 the gluon-gluon fusion $gg \to H$ is the dominant
inclusive Higgs production mechanism at LHC conditions. In this process,
the Higgs production occurs via triangle heavy (top) quark loop.

 It is obvious that the gluon-gluon fusion contribution to the Higgs 
production
at LHC is strongly dependend on the gluon density $xG(x,\mu^2)$ in a
proton. Usually gluon density are described by the 
Dokshitzer-Gribov-Lipatov-Altarelli-Parizi (DGLAP)
evolution equation, where large logarithmic terms proportional to 
$\ln \mu^2$ are taken into
account. However, at the LHC energies typical
values of the incident gluon momentum fractions $x \sim m_H/\sqrt s \sim 
0.008$ (for Higgs boson mass
$m_H = 120$ GeV) are small, and another large logarithmic terms 
proportional to $\ln 1/x$ become important.
These contributions can be taken into account using 
Balitsky-Fadin-Kuraev-Lipatov (BFKL)
evolution equation.
Just as for DGLAP, in this way it is possible to factorize
an observable into a convolution of process-dependent hard matrix 
elements with universal gluon distributions. But as the virtualities 
(and transverse
momenta) of the propagating gluons are no longer ordered, the matrix
elements have to be taken off-shell and the convolution made also over
transverse momentum ${\mathbf k}_T$ with the unintegrated 
($k_T$-dependent) gluon  distribution ${\cal F}(x,{\mathbf k}_T^2)$.
This generalized factorization is called $k_T$-factorization.\\
In the collinear factorization,
the calculation of Higgs production processes is quite complicated even 
at lowest order
because of the heavy quark loops contribution. For example, in Higgs + 
one jet production, triangle and box loops
 occur, and in Higgs + two jet production the pentagon loops occur 
(see~\cite{DD} and references therein). 
However,
the calculations of the Higgs production rates can be simplified in the 
limit of large
top quark mass $m_t \to \infty$~\cite{EGN}. In this approximation the 
coupling of the gluons to the Higgs
via top-quark loop can be replaced by an effective coupling. Thus it 
reduces the number of loops in a given diagram by one. The large $m_t$ 
approximation is
valid to an accuracy of $\sim 5$\% in the intermediate Higgs mass range 
$m_H < 2 m_t$, as long as transverse momenta of the Higgs or final jets 
are smaller than of the top quark mass ($p_T < m_t$).

A particularly interesting quantity is the transverse momentum
distribution of the produced Higgs boson. 
It is well-known that the fixed-order perturbative QCD is applicable
when the Higgs transverse momentum is comparable to the $m_H$. Hovewer, 
the main part of the
events is expected in the small-$p_T$ region ($p_T \ll m_H$), where the 
coefficients
of the perturbative series in $\alpha_s$ are enhanced by powers of large
logarithmic terms proportional to $\ln m_H^2/p_T^2$. Therefore reliable 
predictions at small $p_T$
can only be obtained if these terms will be resummed to all orders.
Recently it was shown~\cite{GK} that in the framework of 
$k_T$-factorization 
approach the soft
gluon resummation formulas are the result of the approximate treatment 
of the solutions of the CCFM  evolution equation (in the 
$b$-representation).

There are several additional motivations for our study of
the Higgs production in the $k_T$-factorization approach. First of all, 
in the standard collinear approach, when the transverse momentum of
the initial gluons is neglected, the transerse momentum of the final 
Higgs boson in $gg \to H$ subprocess is zero. Therefore it is necessary 
to include an initial-state QCD radiation to generate the $p_T$
distributions. In the $k_T$-factorization approach  the 
underlying  partonic subprocess is $gg \to H$ and the 
$k_T$-factorization naturally includes a large part of
 the high-order  perturbative QCD corrections~\cite{RSS}.
Since the upper gluon in the parton ladder is not included in the hard 
interaction, its
transverse momentum is now determined by the properties of the evolution 
equation only.
It means that in the $k_T$-factorization approach the study  of 
transverse momenta distributions in the Higgs production
via gluon-gluon fusion will be direct probe of the unintegrated gluon 
distributions in a proton. In this case
the transverse momentum of the produced Higgs should be equal to the sum 
of the of the initial gluons. Therefore future experimental studies at 
LHC can be used as further test of the non-collinear parton evolution.

In the previous studies~\cite{GK,J1,WMR} the $k_T$-factorization 
formalism 
was applied to calculate transverse momentum distribution of the 
inclusive Higgs production.
The calculations~\cite{GK,WMR} were done using the on-mass shell 
(independent
from the gluon $k_T$) matrix element of the $gg\to H$ subprocess and 
rather the similar results have been obtained. In Ref.~\cite{J1} in the 
framework of MC generator
CASCADE~\cite{J2} the off-mass-shell matrix element obtained by 
F.~Hautmann~\cite{Haut} has been used with full CCFM evolution.
In our paper~\cite{LZ} we have investigated Higgs production at hadron 
colliders
using the full CCFM-evolved unintegrated gluon densities.
Here in order to  illustrate the fact that in the 
$k_T$-factorization approach
the main features of collinear higher-order pQCD corrections are taken 
into account effectively, we give theoretical
predictions for the Higgs + one jet and Higgs + two jet production 
processes.

\section{Basic formulas}
The effective Lagrangian for the Higgs boson coupling to
gluons~\cite{DD} is
$$
  {\cal L}_{\rm {eff}} = {\alpha_s \over 12 \pi}\left(G_F \sqrt 
2\right)^{1/2} G_{\mu \nu}^a G^{a\,\mu \nu} H, \eqno (1)
$$
\noindent
where $G_F$ is the Fermi coupling constant, $G_{\mu \nu}^a$ is the gluon 
field strength tensor
and $H$ is the Higgs field. The triangle vertex $T^{\mu \nu}(k_1,k_2)$ 
for two off-shell gluons having four-momenta
$k_1$ and $k_2$ and color indexes $a$ and $b$ respectively, can be
obtained  easily from the Lagrangian (1):
$$
  T^{\mu \nu}(k_1,k_2) = i \delta^{a b} {\alpha_s \over 3\pi} \left(G_F 
\sqrt 2\right)^{1/2} \left[ k_2^{\mu} k_1^{\nu} -
    (k_1 \cdot k_2) g^{\mu \nu} \right]. \eqno (2)
$$
 The  differential cross section of inclusive Higgs production $p\bar p
\to H +X$ in the $k_T-$factorization approach has been calculated in
~\cite{LZ} and can be written as:
$$
  \displaystyle {d\sigma(p \bar p \to H + X)\over dy_H} = \int 
{\alpha_s^2(\mu^2)\over 288 \pi} {G_F \sqrt 2 \over x_1 x_2 m_H^2 s} 
\left[m_H^2 + {\mathbf p}_T^2\right]^2 \cos^2 (\Delta\varphi) \times 
\atop
    \displaystyle \times {\cal A}(x_1,{\mathbf k}_{1T}^2,\mu^2) {\cal 
A}(x_2,{\mathbf k}_{2T}^2,\mu^2) d{\mathbf k}_{1T}^2 d{\mathbf k}_{2T}^2 
{d(\Delta\varphi) \over 2\pi}, \eqno (3)
$$
where ${\cal A}(x,{\mathbf k}_{T}^2,\mu^2)$ is the unintegrated gluon
distribution, $\Delta\varphi$ the azimutal angle between the initial 
gluon 
momenta ${\mathbf k}_{1T}$ and ${\mathbf k}_{2T}$, and the transverse 
momentum of the produced Higgs boson is ${\mathbf p}_{T} =
 {\mathbf k}_{1T} + {\mathbf k}_{2T}$. It should be noted that this 
process is particulary interesting in $k_T-$factorization, as the 
transverse momenta of the gluons are in the same order as their 
longitudinal momenta ($\sim (10$ GeV))~\cite{J1}.  

\section{Numerical results}
\begin{figure*}[ht]
\begin{center}
\includegraphics[width=0.45\linewidth]{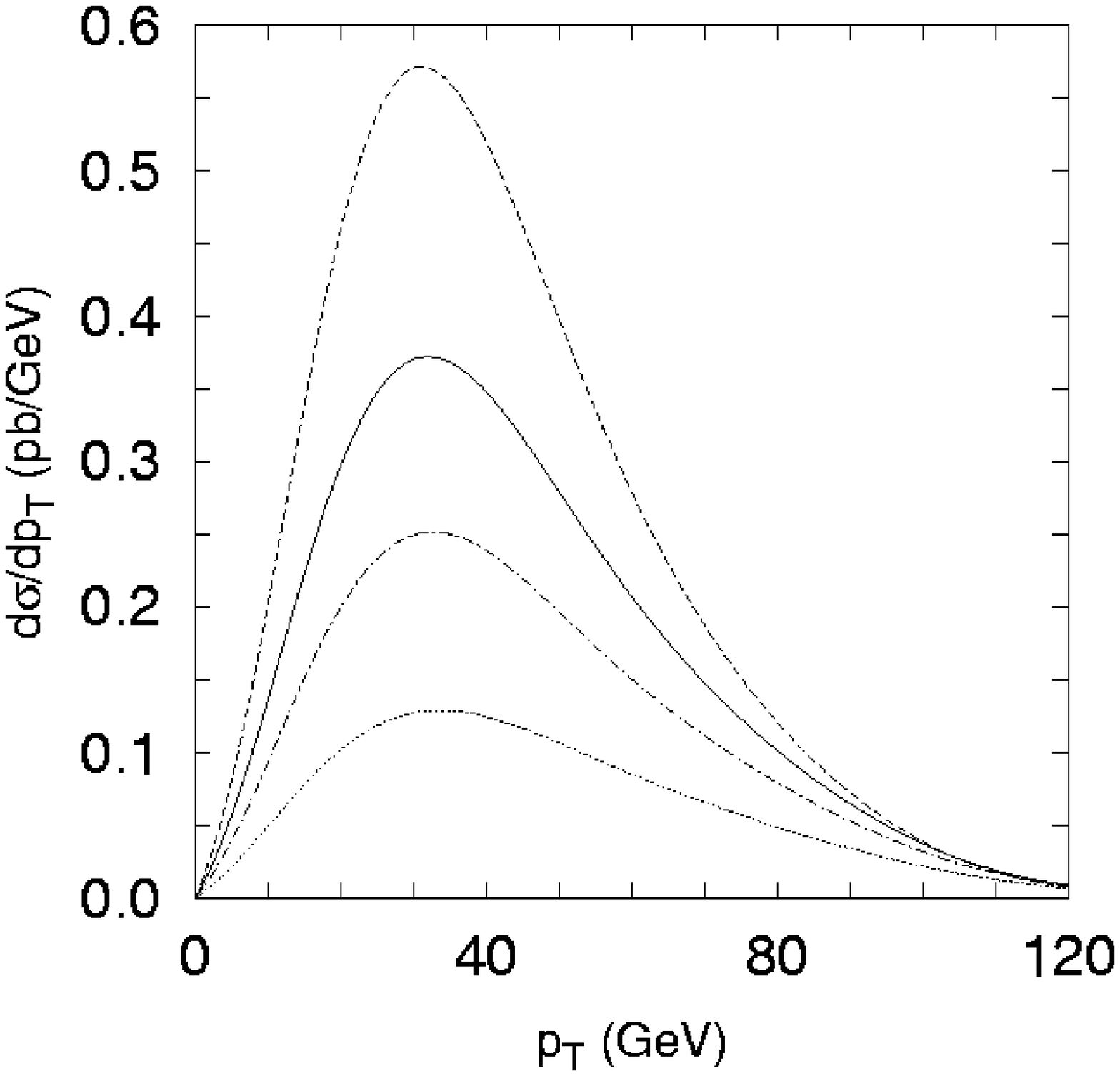}
\includegraphics[width=0.45\linewidth]{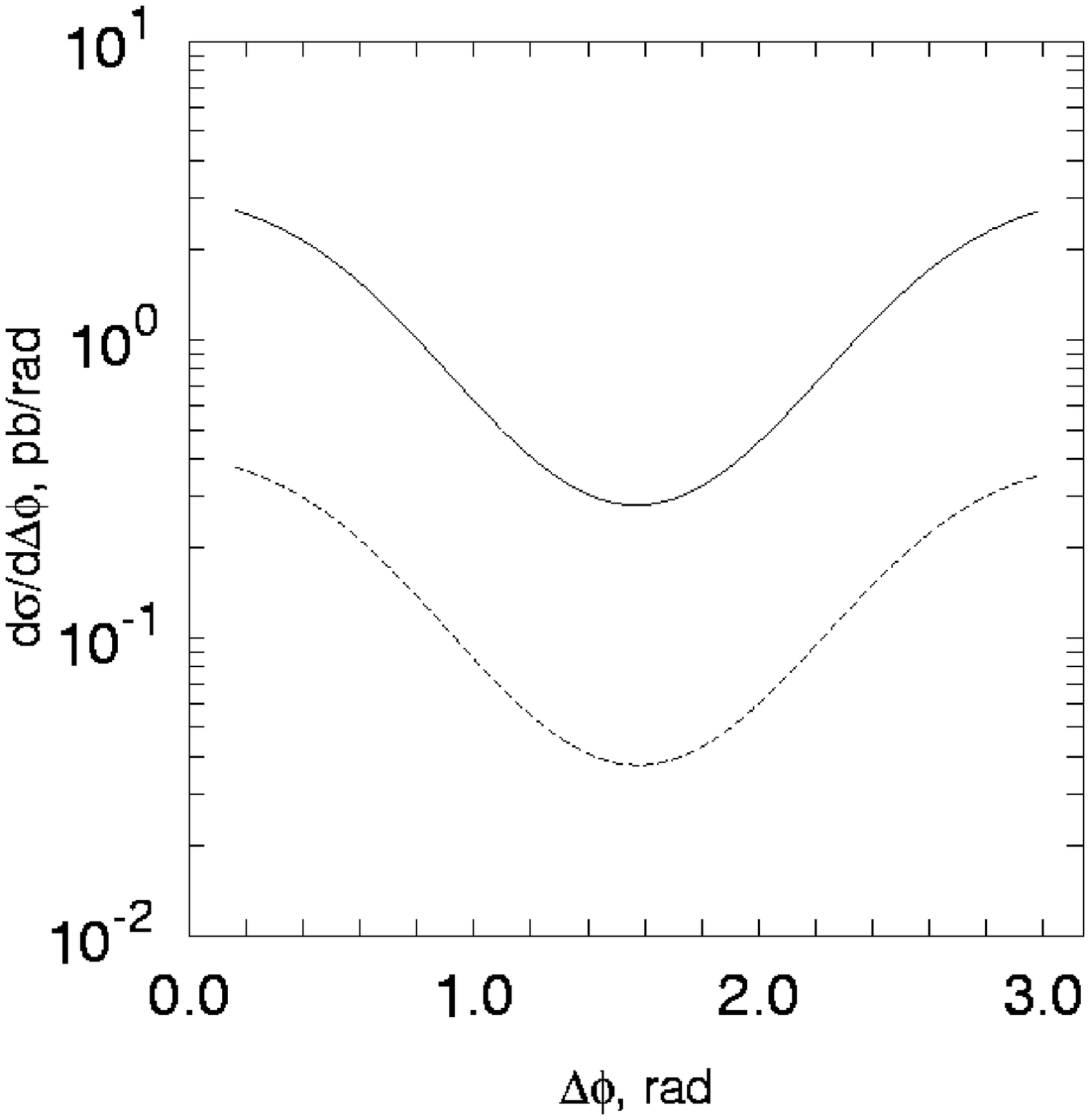}
\end{center}
\vspace*{0.5mm}
\caption{Differential cross section $d\sigma/dp_T$ for 
Higgs + one jet
production at $\sqrt{s} = 14$ TeV: solid, dashed, dash-dotted and
dotted curves correspond to $m_H = 125, 100, 150, 200$ GeV, 
respectively (left column). The jet-jet azimutal angle distribution in 
the Higgs + two jet production: solid and dashed curves correspond to 
the J2003 set 1 and J2003 set 2 unintegrated gluon distributions,
respectively (right column).
\label{fig: higgs}}
\end{figure*}
Our prediction for the total cross secrion and the transverse momentum 
and rapidity distributions
of the inclusive Higgs production at the LHC ($\sqrt s = 14$ TeV)
were done in Ref.~\cite{LZ}.
The calculations were fulfiled for four choices of the Higgs boson mass
($m_H = 100, 125, 150, 200$ GeV) 
under interest in the Standard Model with default scale $\mu^2 = m_H^2$.
We 
used  LO formula for the strong coupling constant $\alpha_s(\mu^2)$ with 
$n_f = 4$ active quark flavours
and $\Lambda_{\rm QCD} = 200$ MeV, such that $\alpha_s(M_Z^2) = 0.1232$.

Here we illustrate how $k_T$-factorization approach can be
used to calculate the semi-inclusive Higgs production rates.
We  choose the one carrying the largest
transverse momentum, and then compute Higgs with an associated
jet cross sections at the LHC energy. We have
applied the usual cut on the final jet transverse momentum
$|{\mathbf p}_{{\rm jet}\,T}| > 20$ GeV. Our predictions for the
transverse momentum distribution of the Higgs + one jet production are
shown in Fig. 1 (left column). One can see the shift of the peak 
position in the
$p_T$ distributions in comparison with inclusive production, which is 
direct consequence of the $|{\mathbf p}_{{\rm jet}\,T}| > 20$ GeV 
cut. 
To demonstrate the possibilities of the $k_T$-factorization approach,
 we calculate azimuthal angle distributions between the two final jet 
transverse momenta in the Higgs + two jet production process, where the 
kinematical cut $|{\mathbf p}_{{\rm jet}\,T}| > 20$ GeV was applied for 
both final jets.  Studing  of these quantities are important to clean
separation of weak-boson fusion and gluon-gluon fusion contributions.
Our results are shown in Fig. 1 (right column), where a 
dip at 90 degrees is seen. This dip at $\Delta\phi\simeq\pi /2$
comes from the $cos(\Delta\varphi)$
in eq. (3). In the approach presented here, the $k_T$ of the initial
gluons is approximately compensated by the transverse momenta of the 
jets~\cite{BZ}: ${\mathbf k}_{T}\simeq{\mathbf p}_{{\rm jet}\,T}$ and, 
consequently, $\Delta\phi\simeq\Delta\varphi$. This dip is 
characteristic
for the loop-induced Higgs coupling to gluons in the framework of
fixed-order pQCD calculations~\cite{HK}. Thus, we illustrate that the 
features
usually interpreted as NNLO effects are reproduced in the 
$k_T-$factorizatiuon with LO matrix elements.
 
 However, we see a very large difference between the predictions based 
on the J2003 gluon densities set 1 and set 2~\cite{J1}, showing the
sensitivity to the shape of the unintegrated gluon distribution.

\section{Conclusions}
 The predictions in the $k_T-$factorization approach are very close
to NNLO pQCD results for the inclusive Higgs
production at the LHC, since the main part of high-order collinear pQCD
corrections is already included in the $k_T-$factorization. In the 
$k_T-$factorization approach the calculation of the associated Higgs+
jets production is much simpler than in the collinear factorization 
approach. However, the large scale dependence of our calculations
(of the order of 20 - 50$\%$) probably indicates the sensitivity to the
 unintegrated gluon distributions.
\section*{Acknowledgments}
N.Z. thanks J. Tran Thanh Van and B. Nicolescu for financial support
and hostly and friendly atmosphere during the Workshop.

\end{document}